\documentclass[aps,prl,twocolumn,showpacs,superscriptaddress,floats]{revtex4-1}
\usepackage[dvips]{graphicx}
\usepackage{amssymb}
\usepackage{amsmath}
\usepackage{url}
\usepackage{amsfonts}
\usepackage{color}
\usepackage{dcolumn}
\usepackage{bm}
\usepackage{pstricks}
\usepackage[dvips,hyperfootnotes=false,breaklinks=true]{hyperref}
\hypersetup{
bookmarks=true,
colorlinks=true,
linkcolor=blue,
anchorcolor=blue,
citecolor=blue,
filecolor=blue,
urlcolor=blue,
bookmarksnumbered=true,
}

\newcommand{\BFCA}{Ba(Fe$_{1-x}$Co$_x$)$_2$As$_2$}
\newcommand{\BFA}{BaFe$_2$As$_2$}

\newcommand{\eps}{\varepsilon}
\newcommand{\vk}{{\mathbf{k}}}
\newcommand{\vR}{{\mathbf{R}}}
\newcommand{\vt}{{\mathbf{\tau}}}
\newcommand{\vQ}{{\mathbf{Q}}}
\newcommand{\vkQ}{{\mathbf{k+Q}}}

\newcommand{\vq}{\mathbf{q}}
\newcommand{\mT}{{\hat{\mathrm{T}}}}
\newcommand{\mG}{{\hat{\mathrm{G}}}}
\newcommand{\mF}{{\hat{\mathrm{F}}}}
\newcommand{\mP}{{\hat{\mathrm{P}}}}

\newcommand{\ut}[1]{\mathrm{\; #1}}
\newcommand{\comment}[1]{}
\newcommand{\ham}{{\mathcal{H}}}

\renewcommand{\Im}{\operatorname{Im}}

 \graphicspath{{./figures/}}

\begin{document}

\title{Effects of Lifshitz Transition on Charge Transport in Magnetic Phases of Fe-Based Superconductors}

\author{Y. Wang}
\affiliation{Department of Physics, University of Florida, Gainesville, Florida 32611, USA}

\author{Maria N. Gastiasoro}
\author{Brian M. Andersen}
\affiliation{Niels Bohr Institute, University of Copenhagen, Universitetsparken 5, DK-2100 Copenhagen,
Denmark}

\author{M. Tomi\'{c}}
\author{Harald O. Jeschke}
\author{Roser Valent\'{i}}
\affiliation{Institut f\"{u}r Theoretische Physik, Goethe-Universit\"{a}t Frankfurt, 60438 Frankfurt am Main,
Germany}
\author{Indranil Paul}
\affiliation{Laboratoire Mat\'{e}riaux et Ph\'{e}nom\`{e}nes Quantiques, Universit\'{e} Paris Diderot-Paris
VII \& CNRS, UMR 7162, 75205 Paris, France}
\author{P. J. Hirschfeld}
\affiliation{Department of Physics, University of Florida, Gainesville, Florida 32611, USA}

\begin{abstract}
The unusual temperature dependence of the resistivity and its in-plane anisotropy observed in the Fe-based
superconducting materials, particularly {\BFCA}, has been a long-standing puzzle. Here, we consider the
effect of impurity scattering on the temperature dependence of the average resistivity within a simple
two-band model of a dirty spin density wave metal. The sharp drop in resistivity below the N\'eel
temperature $T_N$ in the parent compound  can only be understood in terms of a Lifshitz transition
following Fermi surface reconstruction upon magnetic ordering. We show that the observed resistivity
anisotropy in this phase, arising from nematic defect structures, is affected by the Lifshitz transition as
well.
\end{abstract}


\maketitle

Lifshitz transitions (LT) in metals~\cite{lifshitz}, where Fermi surfaces change topology, have mostly been
studied as zero temperature ($T$) phenomena driven by external parameters such as doping and pressure,
etc.~\cite{lifshitz2,buhmann}. Temperature driven LT that can occur in spin or charge density wave phases of
metals have received comparatively less attention. In this context, an interesting aspect of the Fe-based
superconductors (FeSC) is their multiband nature with several hole and electron pockets. After band
reconstruction in the spin density wave (SDW) phase, some of these pockets can disappear due to the increase
of the SDW potential with lowering temperature. Recently, a combined study of electron Raman and Hall
conductivity on SrFe$_2$As$_2$ has reported signatures of such a transition~\cite{gallais2014}. This
motivates us to study the effects of such transitions on the charge transport of the FeSC. Using a model
where current relaxation is due to impurity scattering, we find remarkably strong signatures of such
transitions in both the average resistivity $\rho_{avg}$ and the resistivity anisotropy $\rho_{ani}$ that are
consistent with known experimental trends of these quantities.

The charge transport properties of the FeSC, particularly of BaFe$_2$As$_2$, are currently the subject of
intense research. The $ab$-plane anisotropy of the resistivity $\rho_{ani} \equiv \rho_a - \rho_b$ of the
strain detwinned crystals below the structural transition temperature $T_S$ has an intriguing sign with the
shorter $b$ axis being more resistive than the longer $a$ axis~\cite{tanatar,chu,chu2}. The anisotropy
weakens upon entering the SDW phase even though the magnetic order by itself breaks $C_4$ symmetry.
Furthermore, the anisotropy magnitude in the SDW phase typically increases upon light doping. Together with
other measurements~\cite{kim11,yi11,yi12,zhang12,zhao,nakajima,gallais2013,kasahara,lgreene}, substantial
$\rho_{ani}$ has been taken as strong evidence for intrinsic electronic nematic
behavior~\cite{hu12,fernandes12,chubukov_fernandes_review14}. The behavior of the average resistivity
$\rho_{avg}$, which has received considerably less attention, is also highly unusual~\cite{rullier-albanque}.
In the parent compounds and lightly doped systems, $\rho_{avg}$ falls abruptly below the SDW transition at
$T_N$, in dramatic contrast with conventional SDW systems such as Cr.

Several theoretical works have attempted to explain the origin of $\rho_{ani}$ based on either anisotropic
inelastic scattering with spin fluctuations giving rise to hot spot
physics~\cite{fernandes-schmalian11,Prelovsek,Brydon} or on an anisotropic Drude weight of the
carriers~\cite{dagotto_prl,valenzuela}. Note that, in the 122 systems, where the anisotropy has mostly been
studied, the band structure poses an additional challenge, since the ellipticity of the electron pockets vary
along the $k_z$ axis; the ellipticity at $k_z=0$ and $\pi$ planes have opposite signs~\cite{SuppMat}.
Consequently, in theories where the sign of $\rho_{ani}$ is determined by the ellipticity $\xi_e$ of the
electron pockets on each $k_z$ plane, such as those involving spin fluctuation scattering, at least a partial
cancelation is expected after the $k_z$ average, and the total $\rho_{ani}$ will depend on details of the
band structure.

In contrast, to the best of our knowledge, there is no theory of the characteristic drop in the average
resistivity $\rho_{avg}\equiv (\rho_{a}+\rho_{b})/2$ immediately below $T_N$. Clearly, it is important to
simultaneously account for  this unusual feature of $\rho_{avg}$ in addition to $\rho_{ani}$. A drop in the
inverse Drude weight below $T_N$ has been recovered in simulations \cite{dagotto_prl}  and \textit{ab initio}
calculations~\cite{blomberg}, but this quantity is distinct from the resistivity and includes no information
about the scattering mechanism. Qualitatively, the sharp drop in $\rho_{avg}$ below $T_N$ can be understood
in terms of a collapse in the scattering rate due to the decrease in phase space upon partial gapping of the
Fermi surface, which then overcompensates the loss of carriers. However, since these two competing effects
have the same physical origin, namely the growth of the SDW amplitude with decreasing $T$, the challenge here
is to understand why the scattering rate collapse dominates the resistivity, at least in the undoped and
lightly doped compounds, and whether this collapse is dominated by the elastic or inelastic scattering
channel.

Our focus on impurity scattering can be appreciated from Fig.~1(a), where we fit the resistivity data of
{\BFA} from Ref.~\onlinecite{ishida13} in the high-$T$ paramagnetic phase ($T>T_N\approx 141\ut{K}$) to
$\rho_{avg} = A + BT^2$. We find excellent agreement up to $T\approx 300\ut{K}$, which argues in favor of
\emph{conventional Fermi liquid} and \emph{disorder scattering}, rather than bad-metal
physics~\cite{bad-metal}. More importantly, we find that $A\gg BT_N^2$ by an order of magnitude, implying
that already at $T_N$ the \emph{elastic scattering} from impurities dominates over inelastic processes.

The relevance of impurity scattering to explain $\rho_{ani}$ is currently being debated. Recently, Ishida
\textit{et al.}~\cite{ishida13} reported that, upon annealing, $\rho_{ani}$ of {\BFA} nearly vanished, while
significant anisotropy remained in Co-doped compounds. They argued that $\rho_{ani}$ is due to ``nematogens''
or anisotropic scattering potentials induced by Fe vacancies and Co defects. Such spatially extended defects
aligned preferentially along $a$ direction have also been reported by scanning probe
studies~\cite{chuang10,allan13,grothe12,song11,song12,hanaguri12,zhou11,rosenthal13}. From the theoretical
standpoint, $C_4$ symmetry breaking defect structures around pointlike impurities driven by
orbital~\cite{kontani} or spin~\cite{navarro,navarro_nem} correlations have indeed been found in realistic
models of the Fe-based materials. On the other hand, Kuo and Fisher~\cite{kuo14}, from a comparison of Co and
Ni doped samples, have argued that the strain induced $\rho_{ani}$ does not depend on impurity concentration
and therefore is an intrinsic property of the carriers.

The following are our main results. (i) We show that the characteristic drop in $\rho_{avg} (T)$ in the SDW
phase is a consequence of one or more temperature-driven LT. (ii) The result applies to a multiband system in
a ``dirty'' limit, in which an effective elastic scattering rate $\Gamma > W_0$, where $W_0$ is SDW potential
at $T=0$. In the opposite limit, $\rho_{avg} (T)$ increases in the SDW phase. (iii) Consistent with our
earlier study~\cite{navarro_nem}, we find that extended anisotropic impurity states aligned along $a$
direction give rise to $\rho_{ani} < 0$ in the paramagnetic state. More importantly, we show that the
anisotropy is independent of the ellipticity of the electron pockets provided the scattering is dominantly
intraband. (iv) For parameters relevant for the parent compound, the LT produce a drop in $\rho_{ani}(T)$
below $T_N$ which is consistent with experiments. This feature is suppressed by reducing $W_0$ sufficiently,
which is in qualitative agreement with the measured doping dependence of $\rho_{ani}(T\rightarrow 0)$.

{\it Model.}---We consider the two-band model of Brydon \textit{et al.}~\cite{pBrydon11} along with a mean
field description of the SDW state and introduce intraband impurity scattering. Since our goal is to study
the effect of rapid change of density of states due to a $T$-driven LT, we do not expect orbital physics to
affect the results qualitatively. The Hamiltonian is given by $\ham = \ham_c + \ham_f + \ham_\text{SDW} +
\ham_\text{imp}$. Here, $\ham_c = \sum_{\vk, \sigma}\eps^c_{\vk} c^{\dagger}_{\vk, \sigma} c_{\vk, \sigma}$
and $\ham_f = \sum_{\vk, \sigma} \eps^f_{\vk} f^{\dagger}_{\vk, \sigma} f_{\vk, \sigma}$ describe $c$-hole
and $f$-electron bands, with spin $\sigma$, centered around $\Gamma$ and $X/Y$ points of the 1Fe/cell
Brillouin zone (BZ) with dispersions $\eps_\vk^c = \eps_c + 2t_c(\cos k_x + \cos k_y)$ and $\eps_\vk^f =
\eps_f + t_{f1}\cos k_x \cos k_y - t_{f2}\xi_e(\cos k_x + \cos k_y)$, respectively. $\ham_\text{SDW} =
\sum_{\vk, \sigma} \sigma W c^{\dagger}_{\vk, \sigma} f_{\vk + \vQ, \sigma} + \text{H.c.}$, with $\vQ =
(\pi,0)$. SDW potential $W= W_0 \tanh (2 \sqrt{T_N/T -1})$ for $T \leq T_N$ and zero otherwise. We specify
all energies in  units of $t_c$, and we choose $\eps_c=-3.5$, $\eps_f=3.0$, $t_{f1}=4.0$, $t_{f2}=1.0$,
$T_N=0.04$. Depending on the magnitude of $W_0$, there are either no LT ($W_0<W_e^{\ast}$), or one LT
($W_h^{\ast}>W_0>W_e^{\ast}$) where electron pockets disappear below $T<T_e^{\ast}$, or two transitions
($W_0>W_h^{\ast}$) where, in addition, hole pockets disappear below $T<T_h^{\ast}<T_e^{\ast}$. ($W_e^{\ast},
W_h^{\ast}, T_e^{\ast}, T_h^{\ast} $) depend on the dispersion parameters.

\begin{figure}[!t]
  \includegraphics[width=0.98\linewidth]{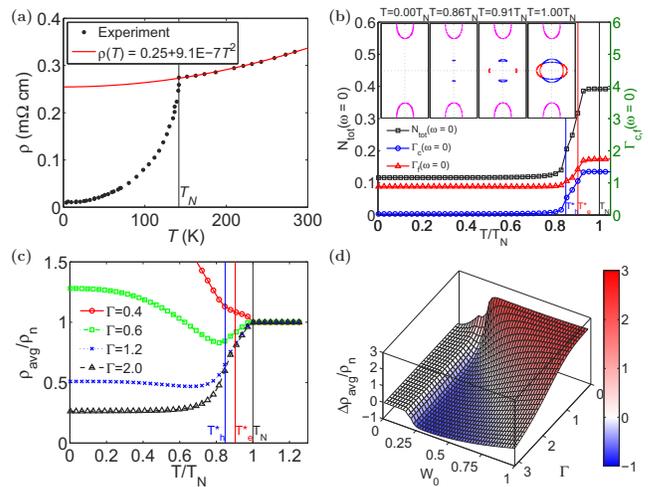}

\caption{(color online). (a) Fit (red line) of resistivity data (black dots)
from Ref.~\onlinecite{ishida13}  in the high-$T$ paramagnetic phase
with $\rho_{avg}=2.5\times10^{-1}+9.1\times10^{-7}T^2$.
(b) $T$ evolution of the total density of states and the $c$- and $f$-electron scattering rates.
Insets: Fermi surface evolution due to $T$ dependence of SDW potential.
$T_N$, $T_e^*$, $T_h^*$ are defined in text.
(c) $T$ dependence of average resistivity for various total scattering rates $\Gamma$.
$\rho_n\equiv \rho_{avg}(T=T_N)$.
(d) $\Delta\rho_{avg}$ (defined in text) dependence on $W_0$ and $\Gamma$.}
\label{fig1}
\end{figure}

The impurity potential $\ham_\text{imp} = \sum_{\vk, \vq, \sigma} V_{\vq} c^{\dagger}_{\vk, \sigma} c_{\vk +
\vq, \sigma} + (c\rightarrow f)$, with $V_{\vq} = V_0 + V_1 (1+2\cos q_x)$, describes scattering of electrons
with both  isotropic pointlike ($V_0$ term) and anisotropic extended impurity ($V_1$ term) potentials. The
latter is modeled by three pointlike scatterers aligned along the long or antiferromagnetic $a$ direction
($x$ axis), and constitutes $T$-independent analogs of the emergent nematogens reported in
Ref.~\onlinecite{navarro_nem}. In the BaFe$_2$As$_2$ system, $V_0$ might represent weak out of plane disorder
not capable of generating nematogens~\cite{navarro,navarro_nem}, and $V_1$ strong in-plane scatterers like Fe
vacancies.

We treat the impurity scattering in the Born approximation, and calculate the $c$ and $f$-scattering rates
$\Gamma^c_{\vk}(\omega) = -\Im [n_i\sum_{\vk^{\prime}} |V_{\vk - \vk^{\prime}}|^2 G_{\vk^{\prime}
\vk^{\prime}}^{c c}(\omega)]$ where $n_i$ is the impurity concentration, and similarly
$\Gamma^f_{\vk}(\omega)$, respectively. We parameterize the two impurity potentials by defining the
scattering rates $\Gamma \equiv n_0 V_0^2 N_\text{tot}$ and $\Gamma_1 \equiv n_1 V_1^2 N_\text{tot}$, where
$(n_0, n_1)$ are the concentrations of pointlike and extended impurities, respectively, and $N_\text{tot}$ is
the total density of states at the chemical potential. Note that, due to $c$-$f$ mixing in the SDW phase, the
Green's functions acquire double indices. Here, $G^{cc}$, $G^{ff}$, etc., denote retarded Green's functions
in the absence of disorder. In other words, we do not calculate the scattering rates self-consistently, but
we checked that doing so does not change the results significantly. We ignore the real parts of these
diagonal (in $c$-$f$ basis) self energies since our aim is only to extract lifetime effects from the impurity
scattering. Similarly, we do not intend to study how impurity scattering affects the SDW potential, and
consequently, we ignore impurity induced off diagonal self energies. We calculate the conductivity in units
of $e^2/\hbar$
\begin{align} \label{eq:conductfreq}
  &\sigma_{ii} =
  -2\sum_{\vk \in \text{BZ}} \int_{-\infty}^{\infty} \frac{d \omega}{\pi}
    \frac{\partial n_\text{F}(\omega)}{\partial \omega}
    \Big\{[2v^c_{\vk,i} \Im\bar{G}^{cc}_{\vk\vk}(\omega)]^2+ \notag \\
  &[v^f_{\vk,i}\Im\bar{G}^{ff}_{\vk\vk}(\omega)]^2
   +4v^c_{\vk,i} v^f_{\vkQ,i}[\Im\bar{G}^{cf}_{\vk\vkQ}(\omega)]^2 \Big\},
\end{align}
where $\bar{G}$ represent the impurity dressed Green's functions, $v_{\vk}^{c,f}$ the velocity vectors, and
$i$ is the $(a,b)$ component of the conductivity tensor $\overleftrightarrow{\sigma}$ (which is diagonal by
symmetry). The factor 2 before $v_{\vk}^{c}$ in the brackets accounts for two hole pockets at
$\Gamma$.

Note that $\rho_{avg}(T)$ and $\rho_{ani}(T)$ are $T$ independent in the paramagnetic phase  of this model,
while the main $T$ dependence in the SDW phase is due to that of the potential $W(T)$. By contrast, in
experiment $\rho_{ani}$ is peaked near $T_N$~\cite{ishida13}. In Ref.~\onlinecite{navarro_nem}, we argued
that this $T$-dependent anisotropy is intimately related to the unusual nature of the nematogens, whereby
they grow in size as the system approaches $T_N$. In the current Letter of the effects of LT, we ignore this
$T$ dependence for simplicity.

{\it Average resistivity.}---We compute first $\rho_{avg}(T)$ by considering only pointlike impurities ($V_1
=0$). In this case, changing the sign of the ellipticity $\xi_e \rightarrow - \xi_e$ is approximately
equivalent to $\rho_a \leftrightarrow \rho_b$ so $\rho_{avg}(T)$ is unchanged (see below). Thus, we compute
it reliably for a given ellipticity, which we fix to $\xi_e=2$. In Figs.~\ref{fig1}(b)--\ref{fig1}(c) we take
$W_0=0.32$ with $W_0/T_N =8$ (consistent with optical measurements~\cite{deGiorgi,Hu2008}), such that $W_0 >
W_h^{\ast}$. The Fermi surface reconstructions associated with the two LT as a function of $T$ are shown in
the inset of Fig.~\ref{fig1}(b). The main panel of (b) shows rapid drops in $N_\text{tot}(\omega=0)$ and in
the scattering rates $\Gamma_{c,f}(\omega=0)$, which is expected from the loss of Fermi surface sheets
associated with the LT. These two competing trends define a crossover in the $T$ dependence of
$\rho_{avg}(T)$ which is shown in Fig.~\ref{fig1}(c). For small $\Gamma \ll W_0$ (clean limit), the loss of
carriers dominates and the resistivity increases with lowering $T$. But for large $\Gamma \gg W_0$ (dirty
limit), the decrease in the scattering rates dominates, and results in a drop in $\rho_{avg}(T)$ whose
magnitude for $\Gamma=2$ is comparable to that of the parent compounds. Note that this scenario of enhanced
conductivity due to increased lifetime, as opposed to that due to enhanced Drude weight~\cite{dagotto_prl},
is consistent with optical measurements~\cite{deGiorgi,Hu2008}. Furthermore, at $T=0$ we get
$\Gamma_c(\omega=0)\ll W_0$ [see Fig.~\ref{fig1}(b)], which agrees with optical conductivity measuring the
Drude peak and the spectral weight depletion due to SDW as well-separated features in
frequency~\cite{deGiorgi,Hu2008}, while the remaining $\Gamma_f(\omega=0)$ contributes to a broad background.

Next, we define the net change in average resistivity $\Delta\rho_{avg} \equiv \rho_{avg}(T=0) -
\rho_{avg}(T=T_N)$ and show how it varies with $\Gamma$ and $W_0$ in Fig.~\ref{fig1}(d). For $W_0 <
W_e^{\ast}$, there is no LT and the change is negligible. For $W_0 > W_e^{\ast}$, such that the system
undergoes at least one LT, we see clearly the dirty (where $\Delta \rho_{avg} < 0$) to clean (where $\Delta
\rho_{avg} > 0$) crossover as $\Gamma$ is changed for fixed $W_0$. This implies that $\rho_{avg}(T)$ of
undoped or lightly-doped compounds can be explained by a LT provided $W_e^{\ast} < W_0 < \Gamma$.

\begin{figure}[!t]
  \includegraphics[width=0.9\linewidth]{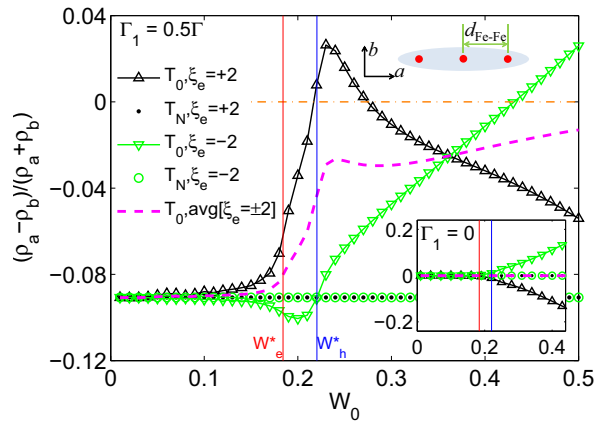}

\caption{(color online). Main panel: $\rho_{ani}$ vs. SDW gap $W_0$ for $\Gamma_1=0.5\Gamma$. Curves for
ellipticity $\xi_e=\pm 2$ at $T=T_0=0$ (upward or downward triangles, respectively) and at $T=T_N$ (dots or
circles, respectively), and for the average of two-plane model (dashed line). The dash-dotted line indicates
$\rho_{ani}=0$. $W_e^{\ast}$, $W_h^{\ast}$ are defined in text. Bottom inset: same quantities for
$\Gamma_1=0$. Top inset: cartoon of extended impurity potential aligned along antiferromagnetic axis $a$.}
\label{fig2}
\end{figure}

{\it Resistivity anisotropy.}---We model the Fermi surface of the 122 systems by calculating the
contributions to the conductivity from the planes $k_z=\pi\;(0)$ with their dispersions differing only in the
$f$-band ellipticities $\xi_e=2\;(-2)$. We calculate the resistivity anisotropy of the planes
$\rho_{ani,\xi_e} \equiv \rho_{a,\xi_e} - \rho_{b,\xi_e}$ separately, and then the experimentally relevant
net anisotropy $\rho_{ani} = {\bar\rho}_a - {\bar \rho}_b$ from the average of the conductivities of the two
planes, i.e., $\bar{\rho}_{i} = \langle\sigma_{i}(k_z)\rangle_{k_z}^{-1} \simeq 2/(\rho_{i, \xi_e}^{-1} +
\rho_{i, -\xi_e}^{-1})$, where $\langle\rangle_{k_z}$ is the exact integral over $k_z$, which we have
approximated by the average of the contributions at $k_z=0$ and $\pi$. As noted earlier for $\Gamma_1=0$,
since $\xi_e \rightarrow - \xi_e$ leads  approximately to $\rho_a \leftrightarrow \rho_b$, the net anisotropy
$\rho_{ani} \simeq 0$ for $T < T_N$, as seen in experiments on annealed samples~\cite{ishida13}, even though
the SDW state itself breaks $C_4$ symmetry (see Fig.~\ref{fig2} bottom inset). The real BaFe$_2$As$_2$ Fermi
surface is considerably more complicated, and there is no exact cancellation between the contributions of
$k_z=0$ and $\pi$ to $\rho_{ani}$, but the true $\rho_{ani}$ will nevertheless be considerably reduced due to
$k_z$ averaging.

We now consider nematogen scattering  by setting  $\Gamma_1 = 0.5 \Gamma$, and calculate the anisotropies
both in the paramagnetic and the SDW phases. Figure~\ref{fig2} shows $\rho_{ani, \xi_e}$ and $\rho_{ani}$ at
$T= T_N$ and $0$ for a wide range of $W_0$. We note that both $\rho_{ani}(T_N) <0$ and $\rho_{ani}(0)<0$,
consistent with experiments. The physical implication of the negative sign is that the nematogens, being
aligned along the $a$ direction, scatter more carriers moving along $b$ than those moving along $a$.
Consequently, we expect this feature to hold even in the presence of interband impurity scattering. Next, we
note that $\rho_{ani,\xi_e}(T_N)$ is independent of the sign of $\xi_e$, which can be understood as follows.
In the paramagnetic phase, assuming intraband-only scattering, the $c$- and $f$-bands decouple. Consequently,
shifting only the $f$ band by $(\pi,\pi)$, keeping the $c$ band unshifted, is an allowed unitary
transformation. $\rho_{ani, \xi_e}(T \geq T_N)$ is invariant under this transformation mapping $\xi_e
\rightarrow -\xi_e$ and is thus  independent of the sign of $\xi_e$.

\begin{figure}[!t]
  \includegraphics[width=0.9\linewidth]{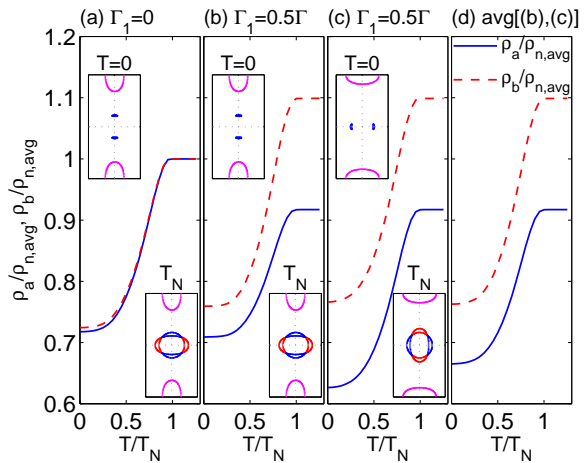}
\caption{(color online). (a) Resistivity $\rho_{a,b}$ at $k_z=\pi$ for a two-plane model with isotropic
scatterers. (b)--(c) Same at $k_z=\pi,0$, respectively, for a two-plane model with anisotropic
scatterers. (d) Average of (b) and (c).}
  \label{fig3}
\end{figure}

Strictly speaking, this argument is invalid in the SDW phase due to $c$-$f$ mixing. Nevertheless for $W_0\ll
W_e^{\ast}$ (relevant for sufficiently doped systems), i.e., without any LT, the Fermi surface reconstruction
is rather weak and we find that $\rho_{ani,\xi_e}(0)$ is practically independent of the sign of $\xi_e$, and,
moreover, $\rho_{ani,\xi_e}(0) \approx  \rho_{ani}(0) \approx \rho_{ani}(T_N)$. However, for $W_0 >
W_e^{\ast}$, the Fermi surface reconstruction due to the LT is significant, and $\rho_{ani,2}(0)$ and
$\rho_{ani,-2}(0)$ are generally different. On the other hand, the magnitude of the net anisotropy is always
less than that in the paramagnetic state, i.e., $\left| \rho_{ani}(0) \right| < \left| \rho_{ani}(T_N)
\right|$. This is due to loss of $N_\text{tot}(\omega=0)$ accompanying the LT (presumably, the associated
gain in carrier lifetime does not affect $\rho_{ani}$). Thus, the LT scenario is able to explain why the
resistivity anisotropy of the undoped and lightly doped systems decrease  as one goes below $T_N$ in the SDW
phase even though the SDW itself breaks $C_4$ symmetry. Furthermore, for $W_0 < W_h^{\ast}$, $\left|
\rho_{ani}(0) \right|$ increases with decreasing $W_0$, which is consistent with the observation that the
resistivity anisotropy in the SDW phase increases with sufficient doping~\cite{footnote-valenzuela}. Finally,
in Fig.~\ref{fig3} we show the $T$ dependence of $\rho_{ani, \xi_e}(T)$ and $\rho_{ani}(T)$ for $W_0=0.2$
(intermediate doping) with $W_0/T_N = 8$.

{\it Conclusions.}---We studied how $T$-driven Lifshitz transitions, where Fermi pockets disappear due to an
increasing SDW potential, affect the average resistivity $\rho_{avg}$ and its anisotropy $\rho_{ani}$ of FeSC
in the magnetic phase. By fitting experimental data, we argued that the dominant current relaxation mechanism
in these materials is impurity scattering. We considered both pointlike and extended impurity (nematogen)
potentials, and showed that the characteristic drop in $\rho_{avg}(T)$ is due to Lifshitz transitions in a
dirty SDW metal. Next, we showed that the nematogen generated $\rho_{ani}$ has the correct sign, namely the
direction with longer lattice constant is less resistive. Within this model, the anisotropy in the
paramagnetic phase is independent of the sign of the ellipticity of the electron pockets. In the SDW phase,
the above holds approximately when the SDW potential is weak enough. The qualitative physics discussed here
is general enough to be of potential interest for transport in other multiband systems showing density wave
instabilities.

We thank M. Breitkreiz, R. Fernandes, Y. Gallais, J. Schmalian, and C. Timm for useful discussions. P.J.H.
and Y.W. were partially supported by Grant NSF-DMR-1005625, B.M.A. and M.N.G. by Lundbeckfond fellowship
(Grant No. A9318), and M.T., H.O.J. and R.V. were partially supported by the Deutsche Forschungsgemeinschaft
through Grant No. SPP 1458.

\setcounter{figure}{0}

\makeatletter
\renewcommand{\thefigure}{S\@arabic\c@figure}
\makeatother

\begin{widetext}
\section{[Supplemental Information]}

\section{$\text{\BFA}$ Fermi surface by first-principles calculations}

In this supplement, we consider the Fermi surface of magnetically ordered {\BFA} obtained from
first-principles calculations using density functional theory (DFT) in order to provide background and
justification for the simple 2-band Hamiltonian model given in the main text, with dispersions
$\eps^{c,f}_\vk$ defined in the 1-Fe Brillouin zone (BZ). The 122 materials (including {\BFA}) have a
different symmetry than, e.g., the 1111 materials for which this Hamiltonian was developed, which requires
accounting for the significant $k_z$ dependence of the Fermi surface. Specifically, the ellipticity $\xi_e$
of the electron pockets changes sign from $k_z=0$ to $k_z=\pi$, as suggested by the DFT calculations shown
below.  From the perspective of our calculation, $\xi_e$ within a purely 2D model cannot be fixed to one sign
or the other  without losing qualitative features of the electronic structure that are important for the
Fermi surface anisotropy. To deal with this problem within a model framework, we consider the contributions
to the conductivity from two representative planes $k_z=0,\pi$, each calculated from a 2D model Hamiltonian
with different ellipticities in the dispersions $\eps^{f}_\vk$, as a realization of the 3D  Fermi surface in
the paramagnetic state (shown in Fig.~\ref{spfig1}(a) as calculated from FPLO~\cite{koepernik1999} for the
{\BFA} orthorhombic structure at $20\ut{K}$ ambient pressure~\cite{rotter2008}).

\begin{figure}[!h]
  \includegraphics[width=0.9\linewidth]{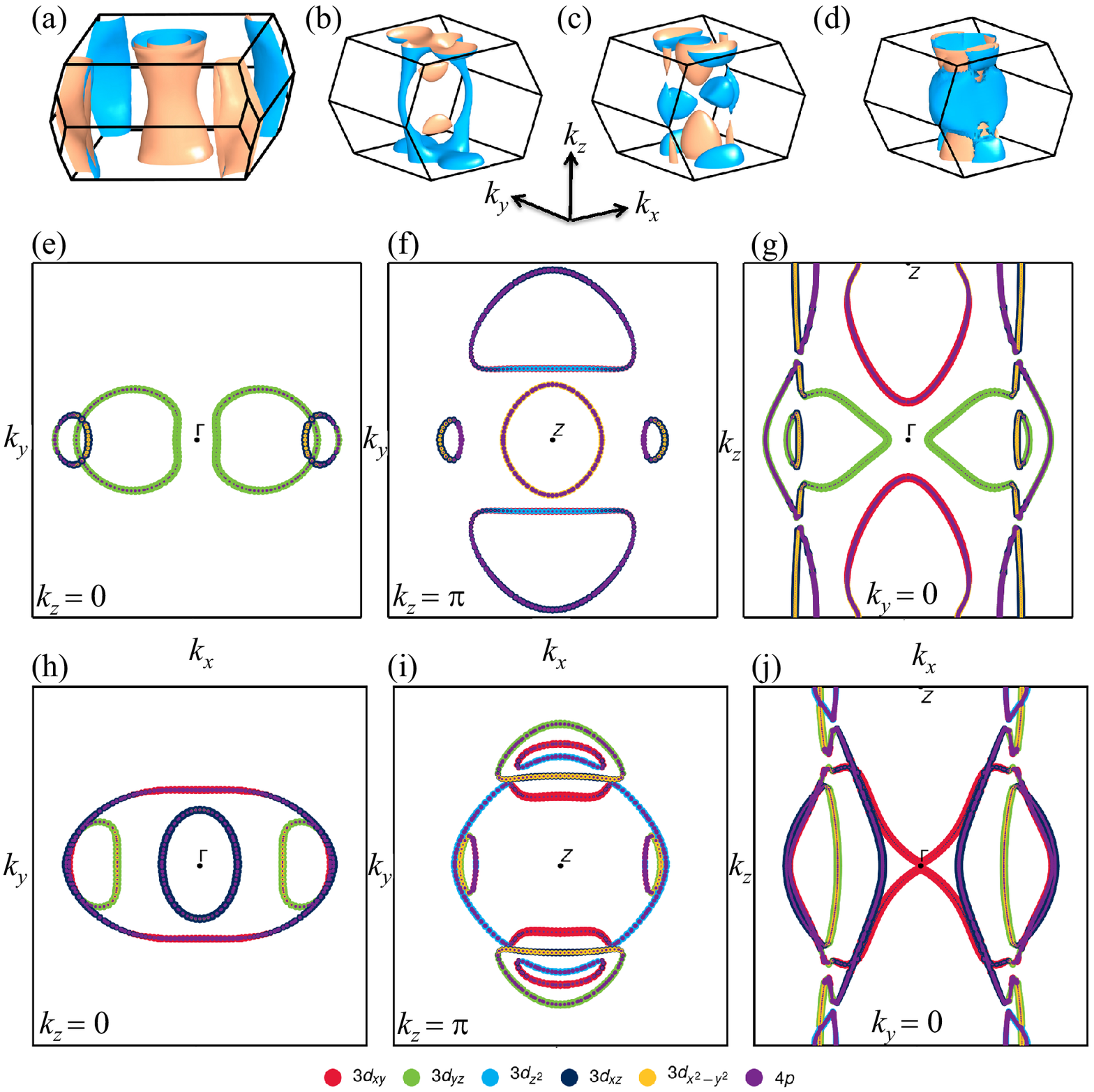}

\caption{Fermi surface of {\BFA} in the paramagnetic phase in the BZ corresponding to the primitive cell of
$Fmmm$ space group (a), in the antiferromagnetic phase in the BZ corresponding to the primitive cell of
$Cccm$ space group with full DFT magnetic moments of $2.1\;\mu_B$ (b), and in the antiferromagnetic phase
with the magnetic moments downscaled to $0.9\;\mu_B$ (c) and to $0.35\;\mu_B$ (d). Second row: cuts of
antiferromagnetic Fermi surface in panel (c) for $k_z=0$ (e), $k_z=\pi$ (f), and $k_y=0$ (g). Third row: cuts
of antiferromagnetic Fermi surface in panel (d) for $k_z=0$ (h), $k_z=\pi$ (i), and $k_y=0$ (j).}
  \label{spfig1}
\end{figure}

Below the N\'{e}el temperature $T_N$, the DFT ground state is known to display the stripelike magnetic order
[$\mathbf{Q}=(\pi,\pi,0)$ in the 2-Fe zone], in agreement with experiment. As shown in Fig.~\ref{spfig1}(b),
the magnetic ground state from the fully self-consistent DFT calculation exhibits a very strong
reconstruction, such that any remnant of the simple magnetic folding shown in the right panel of the
Fig.~\ref{fig1}(b) insert in the main text would be hard to discern. However, this calculation gives an
ordered antiferromagnetic moment of $2.1\;\mu_B$, which is known to exceed the actual ordered moment by a
factor of 2 compared to experiment. Therefore we have performed the same calculation for a restricted moment
corresponding to the experimentally measured value, $0.9\;\mu_B$~\cite{huang2008} and show the obtained 3D
Fermi surface in Fig.~\ref{spfig1}(c) and, additionally, various cuts through this Fermi surface in
Fig.~\ref{spfig1}(e)--(g). It is now clear that the reconstructions in the $k_z=0,\pi$ planes resemble those
one would expect from folding the paramagnetic Fermi surface shown in Fig.~\ref{spfig1}(a), and those in the
spin density wave state from our model in Fig.~\ref{fig1}(b) in the main text. Furthermore, for comparison we
show in Fig.~\ref{spfig1}(d) the 3D Fermi surface corresponding to an ordered moment of $0.35\;\mu_B$ where
the reconstruction is minimal, and in Fig.~\ref{spfig1}(h)--(j) the various cuts through this Fermi surface.

\begin{figure}[!h]
  \includegraphics[width=0.9\linewidth]{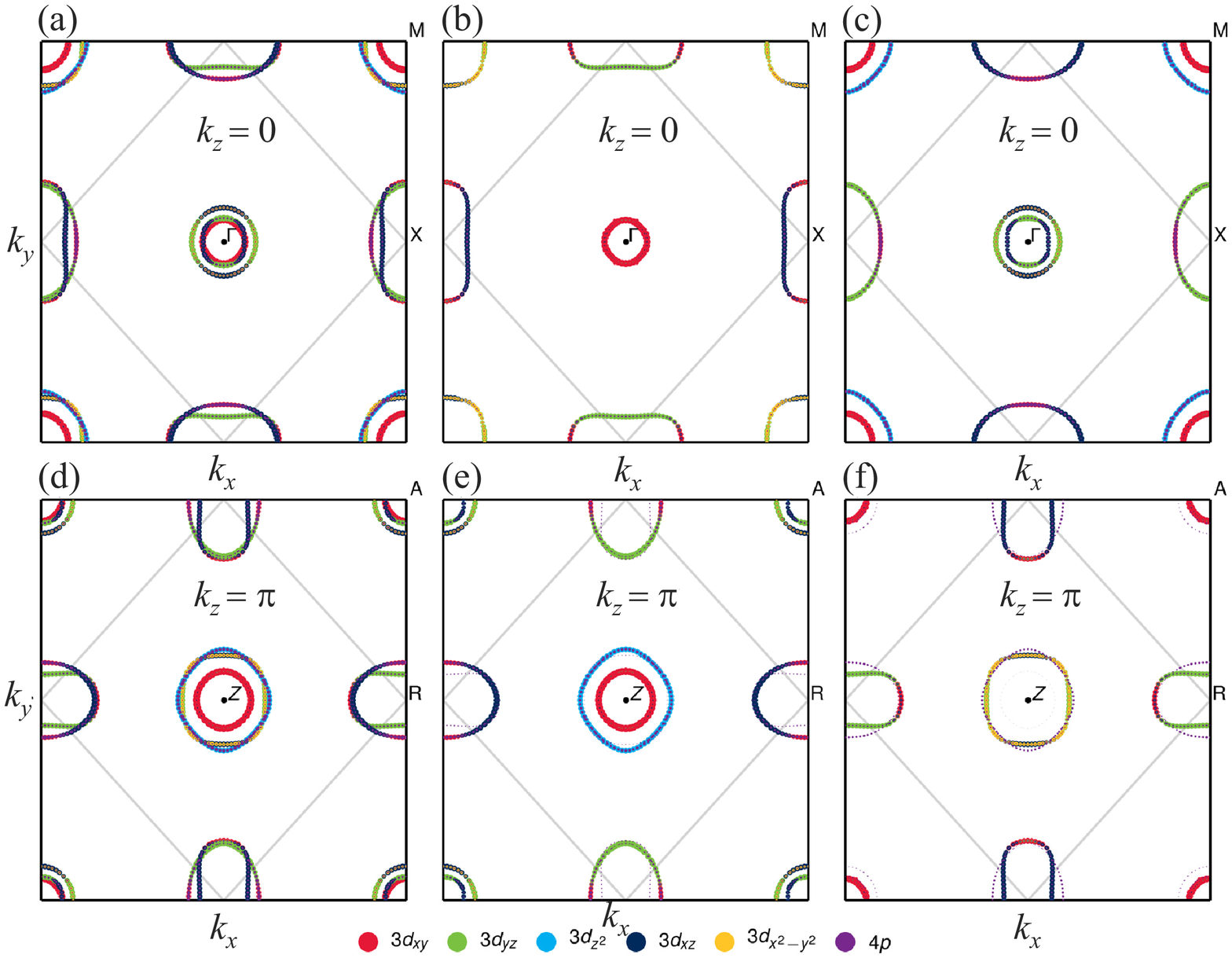}

\caption{\textit{Ab initio} Fermi surface and the unfolded Fermi surface. Top panels: $k_z=0$ cut; lower
panels: $k_z=\pi$ cut. (a, d) \textit{Ab initio} Fermi surface plotted 1Fe zone. The gray line shows the
boundary of 2Fe zone. (b, e) Unfolded Fermi surface obtained by projecting the Bloch states onto the
irreducible subspaces of the glide-mirror group corresponding to the representations $D^{(\vk,+)}$. (c, f) as
in former, only corresponding to $D^{(\vk,-)}$. Note although the unit cell is orthorhombic, we use high
symmetry point labels of the tetragonal unit cell for simplicity. The folding vector $(\pi,\pi,\pi)$ is
apparent when top and bottom panels, e.g., (b) and (f), are compared.}
  \label{spfig2}
\end{figure}

The comparison with the model Fermi surfaces in the main text is not straightforward, however, since
Fig.~\ref{spfig1} corresponds to a 2-Fe unit cell representation in which  the DFT calculations have been
performed, whereas the calculations in the main text from our model use the 1-Fe unit cell. To properly
represent the Fermi surfaces in a 1-Fe model, one should carefully {\it unfold} the
bandstructure~\cite{WeiKu,Andersen2011}, a procedure which is nonintuitive in the complicated 122 crystal
structure. It can be accomplished, however, using symmetries of the crystal in a formal group theoretic
treatment~\cite{tomic2014}. To illustrate our method, we show the \textit{ab initio} Fermi surface
[Fig.~\ref{spfig2}(a, d)] obtained from VASP~\cite{vasp} and unfolded Fermi surface [Fig.~\ref{spfig2}(b, c,
e, f)] at $k_z=0$ and $k_z=\pi$ in Fig.~\ref{spfig2}(a)--(c) and Fig.~\ref{spfig2}(d)--(f), respectively. The
\textit{ab initio} Fermi surface corresponds to the primitive cell containing two translationally
inequivalent Fe atoms, while the unfolded Fermi surface corresponds to the primitive cell containing one Fe
atom. Since the crystal has the glide-mirror symmetry which dictates a degeneracy in the bandstructure, it is
possible to divide the energy bands ($\eps_{n,\vk}$) unambiguously into two sectors ($\eps_{+n,\vk}$ and
$\eps_{-n,\vk}$ defined in 1-Fe zone) that are connected by a folding vector $\vk_f=(\pi,\pi,\pi)$ (i.e.,
$\eps_{+n,\vk}$ = $\eps_{-n,\vk+\vk_f}$). (Note $\eps_{n,\vk}=\eps_{n,\vk+\vk_f}$ is trivial since $\vk_f$ is
the reciprocal lattice vector of 2-Fe zone; it is the division of the $\eps_{n,\vk}$ in 2-Fe zone into
$\eps_{\pm n,\vk}$ in 1Fe zone that is non-trivial and a manifestation of the glide-mirror symmetry.)

This opens up the possibility to formulate a simpler 1-Fe effective model with energy bands either
$\eps_{+n,\vk}$ or $\eps_{-n,\vk}$, whose Fermi surface, when folded along $\vk_f$, results in the Fermi
surface shown in Fig.~\ref{spfig2}(a) and Fig.~\ref{spfig2}(d). In order to identify which portions of the
Fermi surface correspond to the 1-Fe effective model we make use of group theory~\cite{tomic2014}.
Geometrically, we need to map the two translationally inequivalent irons onto each other by means of the
glide-mirror operation and then construct the Bloch-like states adapted to the glide-mirror symmetry.
Formally, this is achieved by projecting  the Bloch-like states onto the irreducible subspaces of the
glide-mirror group, which is just the translation group of the primitive cell of 122 materials extended by
the glide-mirror operation. This extra operation reduces the number of independent degrees of freedom in the
ab-initio electronic structure by a factor of two, allowing us to deduce the Fermi surface of the 1-Fe model.
Specifically, the glide-mirror group has two one-dimensional irreducible representations on the $k_z=0$ and
$k_z=\pi$ planes ($k_z=-\pi$ is equivalent to $k_z=\pi$), given by $D^{(\vk,\pm)}(\mT_\vR)=e^{-i\vk\cdot\vR}$
and $D^{(\vk,\pm)}(\mG)=\pm e^{-i\vk\cdot\vt}$, where $\mT_\vR$ is a translation by lattice vector $\vR$ and
$\mG$ is the glide-mirror operation, with $\vt$  the vector connecting the two translationally inequivalent
iron atoms. The corresponding glide-mirror symmetry adapted Bloch-like basis is generated by the action of
projectors $\mP^{(\vk,\pm)}=\sum_{\mF}D^{(\vk,\pm)\dagger}(\mF)\mF$, where $\mF$ is an operation from the
glide-mirror group. The irreducible representations $D^{(\vk,+)}$ and $D^{(\vk,-)}$ each contain one half of
the bands from the two iron primitive cell and are related by the folding vector $\vk_f$, i.e., $D^{(\vk,-)}
= D^{(\vk+\vk_f,+)}$, which means that each of the irreducible representations $D^{(\vk,\pm)}$ represents one
1-Fe model with the needed folding vector. When the Fermi surface is unfolded according to this prescription,
the ellipticity of the overlapping electron pockets can be unambiguously resolved and the unfolded Fermi
surface shown in either Fig.~\ref{spfig2}(b, e) [from $D^{(\vk,+)}$] or Fig.~\ref{spfig2}(c, f) [from
$D^{(\vk,-)}$] indicates the \emph{sign changing of the ellipticity of the electron pocket from $k_z=0$ to
$k_z=\pi$} as we have assumed in our model [corresponding to Fig.~\ref{spfig2}(c, f)]. The realistic
reconstructed Fermi surface, e.g., Fig.~\ref{spfig1}(c), is obviously considerably more complex than our
two-plane model, but this model nevertheless captures  the main qualitative point  that the resistivity
anisotropy due to isotropic point scatters will be generally considerably  reduced by integration over the
full 122 Fermi surface.

\end{widetext}


\begin{thebibliography}{00}

\bibitem{lifshitz}%
I. M. Lifshitz, Sov. Phys. JETP \textbf{11}, 1130 (1960).

\bibitem{lifshitz2}%
In the context of FeAs, see, e.g., C. Liu \emph{et al.} Phys. Rev. B {\bf 84}, 020509(R) (2011); K. Quader
and M. Widom, Phys. Rev. B \textbf{90}, 144512 (2014).

\bibitem{buhmann}%
For a transport study, see, e.g., J. M. Buhmann and M. Sigrist, Phys. Rev. B  \textbf{88}, 115128 (2013).


\bibitem{gallais2014}%
Y.-X. Yang, Y. Gallais, F. Rullier-Albenque, M.-A. Measson, M. Cazayous, A. Sacuto, J. Shi, D. Colson,
and A. Forget, Phys. Rev. B \textbf{89}, 125130 (2014).

\bibitem{tanatar}%
M. A. Tanatar, A. Kreyssig, S. Nandi, N. Ni, S. L. Bud'ko, P. C. Canfield, A. I. Goldman, and R. Prozorov,
Phys. Rev. B \textbf{79}, 180508 (2009).

\bibitem{chu}%
J.-H. Chu, J. G. Analytis, K. De Greve, P. L. McMahon, Z. Islam, Y. Yamamoto, and I. R. Fisher, Science
\textbf{329}, 824 (2010).

\bibitem{chu2}%
J.-H. Chu, H.-H. Kuo, J. G. Analytis, and I. R. Fisher, Science \textbf{337}, 710 (2012).

\bibitem{kim11}%
Y. K. Kim, H. Oh, C. Kim, D. Song, W. Jung, B. Kim, H. J. Choi, C. Kim, B. Lee, S. Khim, H. Kim, K. Kim, J.
Hong, and Y. Kwon, Phys. Rev. B \textbf{83}, 064509 (2011).

\bibitem{yi11}%
M. Yi, D. Lu, J.-H. Chu, J. G. Analytis, A. P. Sorini, A. F. Kemper, B. Moritz, S.-K. Mo, R. G. Moore, M.
Hashimoto, W.-S. Lee, Z. Hussain, T. P. Devereaux, I. R. Fisher, and Z.-X, Shen, Proc. Natl. Acad. Sci.
U.S.A. \textbf{108}, 6878 (2011).

\bibitem{yi12}%
M. Yi, D. H. Lu, R. G. Moore, K. Kihou, C.-H. Lee, A. Iyo, H. Eisaki, T. Yoshida, A. Fujimori, and Z.-X.
Shen, New J. Phys. \textbf{14}, 073019 (2012).

\bibitem{zhang12}%
Y. Zhang, C. He, Z. R. Ye, J. Jiang, F. Chen, M. Xu, Q. Q. Ge, B. P. Xie, J. Wei, M. Aeschlimann, X. Y. Cui,
M. Shi, J. P. Hu, and D. L. Feng, Phys. Rev. B \textbf{85}, 085121 (2012).

\bibitem{zhao}%
J. Zhao, D. T. Adroja, D.-X. Yao, R. Bewley, S. Li, X. F. Wang, G. Wu, X. H. Chen, J. Hu, and P. Dai, Nat.
Phys. \textbf{5}, 555 (2009).

\bibitem{nakajima}%
M. Nakajima, T. Liang, S. Ishida, Y. Tomioka, K. Kihou, C. H. Lee, A. Iyo, H. Eisaki, T. Kakeshita, T. Ito,
and S. Uchida, Proc. Natl. Acad. Sci. U.S.A. \textbf{108}, 12238 (2011).

\bibitem{gallais2013}%
Y. Gallais, R. M. Fernandes, I. Paul, L. Chauviere, Y.-X. Yang, M.-A. Measson, M. Cazayous, A. Sacuto, D.
Colson, and A. Forget, Phys. Rev. Lett. \textbf{111}, 267001 (2013).

\bibitem{kasahara}%
S. Kasahara, H. J. Shi, K. Hashimoto, S. Tonegawa, Y. Mitzukami, T. Shibauchi, K. Sugimoto, T. Fukuda, T.
Terashima, A. H. Nevidomskyy, and Y. Matsuda, Nature (London) \textbf{486}, 382 (2012).

\bibitem{lgreene}%
H. Z. Arham, C. R. Hunt, W. K. Park, J. Gillett, S. D. Das, S. E. Sebastian, Z. J. Xu, J. S. Wen, Z. W. Lin,
Q. Li, G. Gu, A. Thaler, S. Ran, S. L. Bud'ko, P. C. Canfield, D. Y. Chung, M. G. Kanatzidis, and L. H.
Greene, Phys. Rev. B \textbf{85}, 214515 (2012).

\bibitem{hu12}%
J. P. Hu and C. Xu, Physica (Amsterdam) C \textbf{481}, 215 (2012).

\bibitem{fernandes12}%
R. M. Fernandes and J. Schmalian, Supercond. Sci. Technol. \textbf{25}, 084005 (2012).

\bibitem{chubukov_fernandes_review14}%
R. M. Fernandes, A. V. Chubukov, and J. Schmalian, Nat. Phys. \textbf{10}, 97 (2014).

\bibitem{rullier-albanque}%
See, e.g., F. Rullier-Albenque, D. Colson, A. Forget, and H. Alloul, Phys. Rev. Lett. \textbf{103}, 057001
(2009).

\bibitem{fernandes-schmalian11}%
R. M. Fernandes, E. Abrahams, and J. Schmalian, Phys. Rev. Lett. \textbf{107}, 217002 (2011).

\bibitem{Prelovsek}%
P. Prelov\v{s}ek, I. Sega, and T. Tohyama, Phys. Rev. B \textbf{80}, 014517 (2009).

\bibitem{Brydon}%
M. Breitkreiz, P. M. R. Brydon, and C. Timm, Phys. Rev. B \textbf{90}, 121104(R) (2014).

\bibitem{dagotto_prl}%
S. Liang, G. Alvarez, C. \c{S}en, A. Moreo, and E. Dagotto, Phys. Rev. Lett. \textbf{109}, 047001 (2012).

\bibitem{valenzuela}%
B. Valenzuela, E. Bascones, and M. J. Calderon, Phys. Rev. Lett. \textbf{105}, 207202 (2010).

\bibitem{SuppMat}%
See Supplemental Material, which includes
Refs.~\onlinecite{koepernik1999,rotter2008,huang2008,WeiKu,Andersen2011,tomic2014,vasp} at [url] for
\textit{ab initio} Fermi surfaces.

\bibitem{koepernik1999}%
K. Koepernik and H. Eschrig, Phys. Rev. B \textbf{59}, 1743 (1999).

\bibitem{rotter2008}%
M. Rotter, M. Tegel, D. Johrendt, I. Schellenberg, W. Hermes, and R. P\"ottgen, Phys. Rev. B \textbf{78},
020503 (2008).

\bibitem{huang2008}%
Q. Huang, Y. Qiu, W. Bao, M. A. Green, J. W. Lynn, Y. C. Gasparovic, T. Wu, G. Wu, and X. H. Chen, Phys. Rev.
Lett. \textbf{101}, 257003 (2008).

\bibitem{WeiKu}%
C.-H. Lin, T. Berlijn, L. Wang, C.-C. Lee, W.-G. Yin, and W. Ku, Phys. Rev. Lett. \textbf{107}, 257001
(2011).

\bibitem{Andersen2011}%
O. K. Andersen and L. Boeri, Ann. Phys. (Berlin) \textbf{523}, 8 (2011).

\bibitem{tomic2014}%
M. Tomi{\'c}, H. O. Jeschke, R. Valent{\'i}, Phys. Rev. B \textbf{90}, 195121 (2014).

\bibitem{vasp}%
G. Kresse and J. Hafner, Phys. Rev. B \textbf{47}, 558 (1993); G. Kresse and J. Furthm{\"u}ller,
\textit{ibid}. \textbf{54}, 11169 (1996); Comput. Mater. Sci. \textbf{6}, 15 (1996).

\bibitem{blomberg}%
E. C. Blomberg, M. A. Tanatar, R. M. Fernandes, I. I. Mazin, B. Shen, H.-H. Wen, M. D. Johannes, J.
Schmalian, and R. Prozorov, Nat. Commun. \textbf{4}, 1914 (2013).

\bibitem{ishida13}%
S. Ishida, M. Nakajima, T. Liang, K. Kihou, C. H. Lee, A. Iyo, H. Eisaki, T. Kakeshita, Y. Tomioka, T. Ito,
and S. Uchida, Phys. Rev. Lett. \textbf{110}, 207001 (2013); J. Am. Chem. Soc. \textbf{135}, 3158 (2013).

\bibitem{bad-metal}%
See, e.g., K. Haule and G. Kotliar, New J. Phys. {\bf 11}, 025021 (2009); Z. P. Yin, K. Haule, and G.
Kotliar, Nat. Phys. \textbf{7}, 294 (2011).

\bibitem{chuang10}%
T.-M. Chuang, M. P. Allan, J. Lee, Y. Xie, N. Ni, S. L. Bud'ko, G. S. Boebinger, P. C. Canfield, and J. C.
Davis, Science \textbf{327}, 181 (2010).

\bibitem{allan13}%
M. P. Allan, T.-M. Chuang, F. Massee, Y. Xie, N. Ni, S. L. Bud'ko, G. S. Boebinger, Q. Wang, D. S. Dessau, P.
C. Canfield, M. S. Golden, and J. C. Davis, Nat. Phys. \textbf{9}, 220 (2013).

\bibitem{grothe12}%
S. Grothe, S. Chi, P. Dosanjh, R. Liang, W. N. Hardy, S. A. Burke, D. A. Bonn, and Y. Pennec, Phys. Rev. B
\textbf{86}, 174503 (2012).

\bibitem{hanaguri12}%
T. Hanaguri, (private communication).

\bibitem{zhou11}%
X. Zhou, C. Ye, P. Cai, X. Wang, X. Chen, and Y. Wang, Phys. Rev. Lett. \textbf{106}, 087001 (2011).

\bibitem{song11}%
C.-L. Song, Y.-L. Wang, P. Cheng, Y.-P. Jiang, W. Li, T. Zhang, Z. Li, K. He, L. Wang, J.-F. Jia, H.-H. Hung,
C. Wu, X. Ma, X. Chen, and Q.-K. Xue, Science \textbf{332}, 1410 (2011).

\bibitem{song12}%
C.-L. Song, Y.-L. Wang, Y.-P. Jiang, L. Wang, K. He, X. Chen, J. E. Hoffman, X.-C. Ma, and Q.-K. Xue, Phys.
Rev. Lett. \textbf{109}, 137004 (2012).

\bibitem{rosenthal13}%
E. P. Rosenthal, E. F. Andrade, C. J. Arguello, R. M. Fernandes, L. Y. Xing, X. C. Wang, C. Q. Jin, A. J.
Millis, and A. N. Pasupathy, Nat. Phys. \textbf{10}, 225 (2014).

\bibitem{kontani}%
Y. Inoue, Y. Yamakawa, and H. Kontani, Phys. Rev. B \textbf{85}, 224506 (2012).
%
\bibitem{navarro}%
M. N. Gastiasoro, P. J. Hirschfeld, and B. M. Andersen, Phys. Rev. B \textbf{89}, 100502(R) (2014).

\bibitem{navarro_nem}%
M. N. Gastiasoro, I. Paul, Y. Wang, P. J. Hirschfeld, and B. M. Andersen, Phys. Rev. Lett. \textbf{113},
127001 (2014).

\bibitem{kuo14}%
H.-H. Kuo and I. R. Fisher, Phys. Rev. Lett. \textbf{112}, 227001 (2014).

\bibitem{pBrydon11}%
P. M. R. Brydon, J. Schmiedt, and C. Timm, Phys. Rev. B \textbf{84}, 214510 (2011).

\bibitem{deGiorgi}%
A. Lucarelli, A. Dusza, F. Pfuner, P. Lerch, J. G. Analytis, J.-H. Chu, I. R. Fisher, and L. Degiorgi, New J.
Phys. \textbf{12}, 073036 (2010).

\bibitem{Hu2008}%
W. Z. Hu, J. Dong, G. Li, Z. Li, P. Zheng, G. F. Chen, J. L. Luo, and N. L. Wang, Phys. Rev. Lett.
\textbf{101}, 257005 (2008).

\bibitem{footnote-valenzuela}%
A similar trend was noticed in Ref.~\onlinecite{valenzuela}.

\end{thebibliography}
\end{document}